\documentclass{ws-procs9x6}

\usepackage{bm,epsfig,citesort}

\begin{document}

\title{$\lowercase{f}_2$, $\sigma$, and $2\pi$ exchanges in $\rho$ meson
photoproduction}

\author{Yongseok Oh}

\address{Institute of Physics and Applied Physics,
Yonsei University, \\ Seoul 120-749, Korea \\
E-mail: yoh@phya.yonsei.ac.kr}

\author{T.-S. H. Lee}

\address{Physics Division, Argonne National Laboratory, Argonne,
Illinois 60439, U.S.A. \\
E-mail: lee@theory.phy.anl.gov}


\maketitle

\abstracts{
The $\sigma$ meson exchange model for $\rho$ photoproduction at low
energies is re-examined and a new model is developed by considering
explicit two-pion exchange and the $f_2$ tensor meson exchange.
The $f_2$ exchange model, which is motivated by the low energy
proton-proton elastic scattering, is constructed by fully taking into
account of the tensor structure of the $f_2$ meson interactions.
Phenomenological informations together with tensor meson dominance and
vector meson dominance assumptions are used to estimate the $f_2$
meson's coupling constants.
For $2\pi$ exchange, the loop terms including intermediate $\pi N$
and $\omega N$ channels are calculated using the coupling
constants determined from the study of pion photoproduction.
It is found that our model with $f_2$ and $2\pi$ exchanges can
successfully replace the commonly used $\sigma$ exchange model that
suffers from the big uncertainty of the coupling constants.
We found that the two models can be distinguished by examining
the single and double spin asymmetries.}

Recently the measurements on the electromagnetic production of vector
mesons from the nucleon targets have been reported from the CLAS of
TJNAF~\cite{CLAS00-CLAS01,CLAS01b,CLAS03}, GRAAL of Grenoble~\cite{graal},
and LEPS of SPring-8~\cite{leps,mibe}.
More data with high accuracy on various physical quantities of these
processes are expected to come soon.
These new data replace the limited old data of low statistics and provide
an opportunity to understand the production mechanism of vector mesons at
low energies.
They are also expected to shed light on resolving the `missing resonance'
problem~\cite{CR00,OTL01,ZLB98c-Zhao01,TL03}.

However, it is well-known that thorough understanding of the nonresonant
background mechanisms is crucial to extract the properties of the
resonances and to identify any missing resonances from the data for meson
production processes~\cite{OL02,PM02c}.
As a continuation of our effort in this direction~\cite{OTL01,OL02}, we
explore the nonresonant mechanisms of $\rho$ photoproduction in this
work.

Through the analyses of vector meson photoproduction, we learned that
at high energies the total cross sections are dominated by the
Pomeron exchange, which is responsible to the diffractive features of
the data at small $t$.
However, at low energies, meson exchanges or Reggeon exchanges are
dominant over the Pomeron exchange and responsible to the bump structure
of the total cross section near threshold.
In $\omega$ photoproduction, it is well-known that one-pion exchange is
the most dominant process.
In $\rho$ photoproduction, however, the situation is not so clear.
There are, in general, two models for the major production mechanism of
$\rho$ photoproduction at low energies.
One is the $\sigma$ meson exchange model~\cite{FS96,OTL00} and the other
is the $f_2$ meson exchange model~\cite{Lage00,KV01a-KV02,OL03}.

\begin{figure}[t]
\centering
\epsfig{file=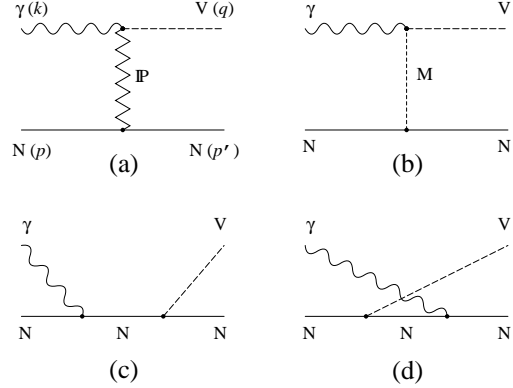, width=0.6\hsize}
\caption{Models for $\rho$ photoproduction. (a,b) $t$-channel Pomeron and
one-meson exchanges ($M=f_2,\pi,\eta,\sigma$). (c,d) $s$- and $u$-channel
nucleon pole terms.}
\label{fig:rho1}
\end{figure}

The $\sigma$ exchange model is motivated \cite{FS96} by the observation
that the $\rho \to \pi \pi \gamma$ decay is much larger than the other
radiative decays of the $\rho$ meson such as $\rho \to \pi \gamma$.
Therefore the role of $2\pi$ exchanges is expected to be important in
the production mechanism of $\rho$ photoproduction.
It is then assumed that the $\pi\pi$ in the $\pi\pi\gamma$ channel can
be modeled as a $\sigma$ meson such that the $\sigma\rho\gamma$ vertex can
be defined for calculating the $\sigma$ exchange mechanism as illustrated
in Fig.~\ref{fig:rho1}(b).
In practice, the product of the coupling constants
$g_{\sigma\rho\gamma} g_{\sigma NN}$ of this tree-diagram is adjusted to
fit the cross section data of $\rho$ photoproduction at low energies.
The parameters of the $\sigma$ exchange model determined in this way
are~\cite{FS96,OTL00}
\begin{eqnarray}
&& M_\sigma = 0.5 \mbox{ GeV}, \qquad g^2_{\sigma NN}/4\pi = 8.0, \qquad
g_{\sigma\rho\gamma}^{} = 3.0.
\label{para:sigma}
\end{eqnarray}
The resulting $\sigma$ mass parameter is close to the value
$M_\sigma= 0.55 \sim 0.66$ GeV of Bonn potential \cite{MHE87}.
If we further take the value $g^2_{\sigma NN}/4\pi = 8.3 \sim 10$ from Bonn
potential, we then find that the resulting $g_{\sigma\rho\gamma}$ is close
to the values from the QCD sum rules,
$ g_{\sigma\rho\gamma}^{}(\mbox{\small QCDSR}) = 3.2 \pm 0.6$
\cite{GY01} or $2.2 \pm 0.4$ \cite{AOS02}.
However such a large value of $g_{\sigma\rho\gamma}^{}$ overestimates
the observed $\rho \to \pi^0 \pi^0 \gamma$ decay width
by two orders of magnitude~\cite{GY01a,BELN01,SND02,OK03}.
If we accept the empirically estimated but model-dependent value of the
SND experiment~\cite{SND02}, $\mbox{BR}(\rho \to \sigma
\gamma) = (1.9 \stackrel{+ 0.9}{\mbox{\scriptsize $-0.8$}} \pm 0.4)
\times 10^{-5}$, we get
\begin{equation}
|g_{\sigma\rho\gamma}| \approx 0.25.
\end{equation}
This value is smaller than that of Eq.~(\ref{para:sigma}) by an order
of magnitude.
Therefore, the $\sigma$ exchange model suffers from the
big uncertainty of $g_{\sigma\rho\gamma}$.
Furthermore, there is yet no clear particle identification of the $\sigma$
meson and the use of $\sigma$ exchange in defining $NN$ potential
has been seriously questioned.
Thus it is possible that the $\sigma$ exchange may not be the right major
mechanism for $\rho$ photoproduction.

\begin{figure}[t]
\centering
\epsfig{file=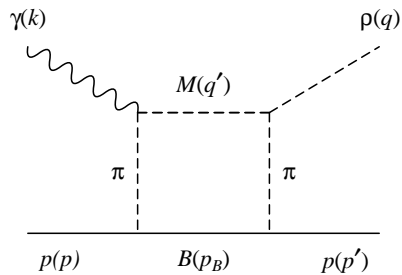, width=0.5\hsize}
\caption{$2\pi$ exchange in $\rho$ photoproduction. The intermediate
meson state ($M$) includes $\pi$ and $\omega$, and the
baryon ($B$) includes the nucleon.}
\label{fig:2pi}
\end{figure}

Therefore, we take a different approach for $\rho$
photoproduction~\cite{OL03} in this work.
Here we consider the $f_2$ exchange and two-pion exchange mechanisms.
Instead of considering the radiative decay of the $\rho$ through the
$\sigma$, we consider the consequences of the strong $\rho^0 \to \pi^+\pi^-$
decay which accounts for almost the entire decay width of the $\rho$ meson.
With the empirical value of the $\rho$ meson decay width, one can define the
$\rho\pi\pi$ vertex which then leads naturally to the two-pion exchange
mechanism illustrated in Fig.~\ref{fig:2pi} with $M=\pi$ in the intermediate
state.
Clearly, this two-pion exchange mechanism is a part of the one-loop
corrections discussed in Ref.~\refcite{OL02} for $\omega$ photoproduction.
A more complete calculation of one-loop corrections to $\rho$
photoproduction is accomplished in this work by including not only the
intermediate $\pi N$ state but also the intermediate $\omega N$ state.

The $f_2$ exchange model for $\rho$ photoproduction was motivated by the
the analyses for $pp$ elastic scattering~\cite{DL92}.
In the study of $pp$ scattering at low energies the secondary Regge
trajectory is important, which is represented by the $f$ trajectory.
The idea of Pomeron-$f$ proportionality then had been used to model the
Pomeron couplings from the $f_2$ couplings~\cite{Freu62-Freu71,CGZ71,KS73}
before the advent of the soft Pomeron model by Donnachie and
Landshoff~\cite{DL84}.
By considering the important role of the $f$ trajectory in $pp$ scattering,
it is natural to consider the $f_2$ exchange model for vector meson
photoproduction.  However, the $f_2$ exchange model developed in
Refs.~\refcite{Lage00,KV01a-KV02} for $\rho$ photoproduction used the
Pomeron-$f$ proportionality in the reverse direction.
Namely, they assume that the structure of the $f_2$ couplings is the same
as that of the soft Pomeron exchange model.
Thus the $f_2$ tensor meson was treated as a $C=+1$ isoscalar photon, i.e.,
a vector particle.
In addition, the fit to the data is achieved by introducing an additional
adjustable parameter to control the strength of the $f_2$
coupling~\cite{Lage00}.
In this work, we elaborate an $f_2$ exchange model starting from effective
Lagrangians constructed by using the empirical information about the tensor
properties of the $f_2$ meson.
The main objective of this work is to construct a model including this newly
constructed $f_2$ exchange amplitude and explicit two-pion exchange
amplitude discussed above.

We now construct an $f_2$ exchange model solely based on the
tensor structure of the $f_2$ meson.
We will use the experimental data associated with the $f_2$ meson, the tensor
meson dominance, and vector meson dominance assumptions to fix the $f_2$
coupling constants~\cite{Renn71,Raman71a,OL03}, such that the strength of
the resulting $f_2$ exchange amplitude is completely fixed in this
investigation.
Following Refs.~\refcite{Gold68,PSMM73}, the effective Lagrangian accounting
for the tensor structure of the $f_2 NN$ interaction reads
\begin{equation}
\mathcal{L}_{fNN} = -2i \frac{G_{fNN}^{}}{M_N} \bar{N} (\gamma_\mu
\partial_\nu + \gamma_\nu \partial_\mu ) N f^{\mu\nu}
+ 4 \frac{F_{fNN}^{}}{M_N} \partial_\mu \bar{N} \partial_\nu N
f^{\mu\nu},
\end{equation}
where $f^{\mu\nu}$ is the $f_2$ meson field and $M_N$ is the nucleon
mass.
The coupling constants were first estimated by using the dispersion relations
in the analyses of the backward $\pi N$ scattering~\cite{Gold68} and the
$\pi\pi\to N\bar{N}$ partial-wave amplitudes.
Here we use~\cite{Hede77-BK77}
\begin{equation}
{G_{fNN}^2}/{4\pi} = 2.2, \qquad F_{fNN} = 0.
\label{eq:cut}
\end{equation}

The most general form for the $fV\gamma$ vertex satisfying gauge
invariance reads~\cite{Renn71}
\begin{equation}
\langle \gamma(k) V(k') | f_2 \rangle = \frac{1}{M_f}
\epsilon^\kappa \epsilon'^\lambda f^{\mu\nu} A_{\kappa \lambda
\mu\nu}^{fV\gamma}(k,k'),
\label{tmd}
\end{equation}  
where $\epsilon$ and $\epsilon'$ are the polarization vectors of the
photon and the vector meson, respectively, and the form of
$A^{fV\gamma}$ can be found in Ref.~\refcite{OL03}.
The tensor meson dominance together with the vector meson dominance
constrain~\cite{Renn71} the coupling constants of $A^{fV\gamma}$.
The details on the $f_2$ interactions and
tensor meson dominance are given in Ref.~\refcite{OL03}.

\begin{figure}[t]
\centering
\epsfig{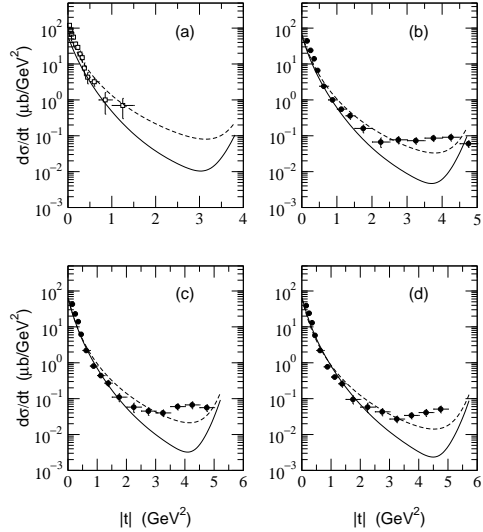}
\caption{Differential cross sections of model (A) and (B) at $E_\gamma =
3.55$ GeV. The dashed line is the result of model (A) and the solid line is
that of model (B).
Experimental data are from Ref.~\protect\refcite{CLAS01b}.}
\label{fig:dsdt}
\end{figure}

\begin{figure}[t]
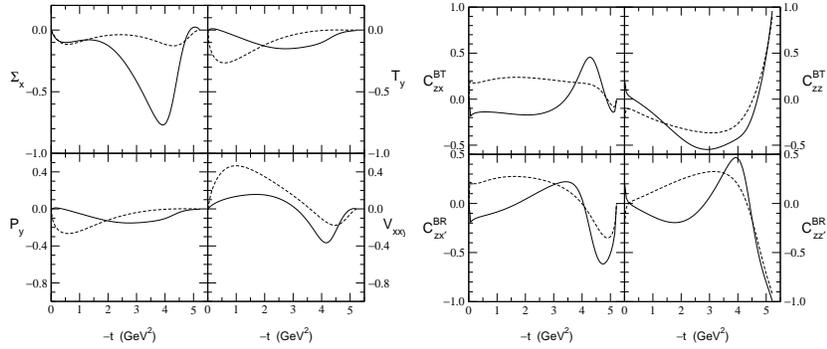

\centering
\epsfig{file=fig4.eps, width=0.47\hsize}
\epsfig{file=fig5.eps, width=0.47\hsize}
\caption{(left panel)
Single spin asymmetries of model (A) and (B) at $E_\gamma =
3.55$ GeV.
(right panel)
Double spin asymmetries $C^{\rm BT}_{zx}$, $C^{\rm BT}_{zz}$,
$C^{\rm BR}_{zx'}$, and $C^{\rm BR}_{zz'}$ of Model (A) and (B) at $E_\gamma =
3.55$ GeV.
Notations are the same as in Fig.~\ref{fig:dsdt}.}
\label{fig:3.55-pols}
\end{figure}

In this work, we explore two models: model (A) includes the Pomeron,
$\sigma$, $\pi$, $\eta$ exchanges, and the $s$- and $u$-channel nucleon
terms, while model (B) is constructed by replacing the $\sigma$
exchange in model (A) by the $f_2$ and $2\pi$ exchanges. (See
Figs.~\ref{fig:rho1} and \ref{fig:2pi}.)
The full calculations of the $\gamma p \rightarrow \rho^0 p$ differential
cross sections from model (A) and (B) are compared in Fig.~\ref{fig:dsdt}.
{}From those figures, one may argue that model (B) is slightly better in
small $t$ region.
However, it would be rather fair to say that the two models are comparable
in reproducing the data.
We therefore explore their differences in predicting
the spin asymmetries, which are defined, e.g., in Ref.~\refcite{TOYM98}.
The results for the single and double spin asymmetries are shown in
Fig.~\ref{fig:3.55-pols} for $E_\gamma = 3.55$ GeV.
Clearly the spin asymmetries would be useful to distinguish the two
models and could be measured at the current experimental facilities.
Of course our predictions are valid mainly in the small $t$ region since
the $N^*$ excitations, which are expected to be important at large $t$
\cite{OTL01}, are not included in this calculation.
Therefore, measurements of such quantities at small $t$ region should be
crucial to understand the main non-resonant production mechanisms
of $\rho$ photoproduction at low energies.

Finally let use mention about the role of the $f_2$ exchange in $\phi$
photoproduction.
In this case, we can consider the exchanges of the $f_2(1270)$ and
$f_2'(1525)$ mesons.
However such exchanges are expected to be negligible {\em if\/} 
the $f_2$ and $f_2'$ mixing is close to the ideal mixing.
This is because the ideal mixing makes the $f_2 \phi \gamma$ and $f_2' NN$
couplings vanish, although $f_2 NN$ and $f_2' \phi\gamma$ do not.
Since the amplitude of this process contains $g^{}_{f_2 \phi\gamma}
G^{}_{f_2 NN}$ or $g^{}_{f_2' \phi\gamma} G^{}_{f_2' NN}$, its contribution
is expected to be small if the $f_2$-$f_2'$ mixing is close to the ideal
mixing.

\section*{Acknowledgments}
Y.O. was supported by Korea Research Foundation Grant
(KRF-2002-015-CP0074) and T.-S.H.L. was supported by
U.S. DOE Nuclear Physics Division Contract No. W-31-109-ENG-38.

\end{document}